**State dependent effects on the frequency response of prestin's real and imaginary components of nonlinear capacitance.**

**Joseph Santos-Sacchi [1, 2, 4], Dhasakumar Navaratnam [1, 2, 3] and Winston Tan [1]**

*[1] Surgery (Otolaryngology), [2] Neuroscience, [3] Neurology and [4] Cellular and Molecular Physiology, Yale University School of Medicine, 333 Cedar Street, New Haven, CT 06510, USA*

***Running title:*** complex NLC is state dependent

***Send correspondence to:***
        Joseph Santos-Sacchi
        Surgery (Otolaryngology), Neuroscience, and Cellular and Molecular Physiology
        Yale University School of Medicine
        BML 224, 333 Cedar Street
        New Haven, CT  06510
        Phone:  (203) 785-7566
        e-mail:  joseph.santos-sacchi@yale.edu

**Acknowledgments:**

This research was supported by NIH-NIDCD R01 DC016318 (JSS) and R01 DC008130 (JSS, DN). Thanks to Kuni Iwasa and Rick Rabbitt for discussions.

**Abstract**

The outer hair cell (OHC) membrane harbors a voltage-dependent protein, prestin (SLC26a5), in high density, whose charge movement is evidenced as a nonlinear capacitance (NLC). NLC is bell-shaped, with its peak occurring at a voltage, $V_h$, where sensor charge is equally distributed across the plasma membrane. Thus, $V_h$ provides information on the conformational state of prestin. $V_h$ is sensitive to membrane tension, shifting to positive voltage as tension increases and is the basis for considering prestin piezoelectric (PZE). NLC can be deconstructed into real and imaginary components that report on charge movements in phase or 90 degrees out of phase with AC voltage. Here we show in membrane macro-patches of the OHC that there is a partial trade-off in the magnitude of real and imaginary components as interrogation frequency increases, as predicted by a recent PZE model (Rabbitt, 2020). However, we find similar behavior in a simple kinetic model of prestin that lacks piezoelectric coupling, the *meno presto* model. At a particular frequency, $F_{is}$, the complex component magnitudes intersect. Using this metric, $F_{is}$, which depends on the frequency response of each complex component, we find that initial $V_h$ influences $F_{is}$; thus, by categorizing patches into groups of different $V_h$, (above and below -30 mV) we find that $F_{is}$ is lower for the negative $V_h$ group. We also find that the effect of membrane tension on complex NLC is dependent, but differentially so, on initial $V_h$. Whereas the negative group exhibits shifts to higher frequencies for increasing tension, the opposite occurs for the positive group. Despite complex component trade-offs, the low-pass roll-off in absolute magnitude of NLC, which varies little with our perturbations and is indicative of diminishing total charge movement, poses a challenge for a role of voltage-driven prestin in cochlear amplification at very high frequencies.

Prestin (SLC26a5) is a membrane-bound voltage-dependent protein (Santos-Sacchi and Dilger, 1988; Zheng et al., 2000), whose activity drives outer hair cell (OHC) electromotility (eM) (Liberman et al., 2002), the latter yet considered to provide for cochlear amplification at very high acoustic frequencies (50 - 100 kHz), despite recent arguments against this concept (Santos-Sacchi and Tan, 2018; Santos-Sacchi et al., 2019; Vavakou et al., 2019; Santos-Sacchi and Tan, 2020). Regardless, the voltage dependent activity of prestin itself has been well characterized over the years (Ashmore, 1990; Santos-Sacchi, 1991), typically analyzed by evaluating membrane capacitance ($\delta Q/\delta V$) (i.e., **N**on**L**inear**C**apacitance, NLC). The proteins' aggregate sensor charge displays a sigmoidal Boltzmann (Q/V) relationship, being quantitatively characterized by $V_h$, the voltage at half maximal charge redistribution across the membrane (near -40 mV); $z$, the apparent unitary charge (near 0.75); and $Q_{max}$ the maximum charge of a given protein population within the membrane. The parameter $z$ is alternatively viewed as a voltage sensitivity, providing a slope factor of the Q/V function (near 35 mV), which indicates a very shallow voltage dependence. Thus, the full range of prestin's voltage responsiveness spans over 300 mV.

Prestin has been deemed a piezoelectric (PZE) protein since Iwasa discovered tension sensitivity of NLC under whole cell (WC) voltage clamp (Iwasa, 1993). Indeed, prestin's piezoelectric coefficient is extraordinary, estimated to be four orders of magnitude greater that man made elements (Dong et al., 2002). Under WC voltage clamp, $V_h$ shifts rightward and peak NLC decreases as membrane tension increases (Iwasa, 1993; Gale and Ashmore, 1994; Kakehata and Santos-Sacchi, 1995). In membrane patches, however, peak NLC (or $Q_{max}$) does not alter despite substantial shifts in $V_h$ (Gale and Ashmore, 1997a; Santos-Sacchi and Tan, 2020). The likely reason why apparent $Q_{max}$ is decreased in WC mode is because tension delivery is inhomogeneous (Gale and Ashmore, 1997a), and differentially impinges on multiple independent micro domains of prestin (Santos-Sacchi, 2002). The disparate summation in WC measures of NLC therefore shows changes in apparent $z$ resulting in a reduced peak NLC. Again, the reduction in total charge with tension is not a characteristic of prestin itself, as membrane patch studies clearly show.

Recently, Rabbitt (Rabbitt, 2020) suggested that the PZE nature of prestin provides power to the cochlear amplifier that correlates with the imaginary component of NLC, that component of charge movement that alters phase in response to viscous load. Such modelled behavior of prestin sensor charge movement, which for a piezo component is coupled to imposed stress, was predicted to account for the apparent incongruity of prestin's low-pass roll-off in the absolute magnitude of complex NLC (Santos-Sacchi and Tan, 2020) and the expected high frequency behavior of prestin required by cochlear modelers to account for high frequency cochlear amplification. Here we thoroughly evaluate the real and imaginary components of NLC in OHC macro-patches and find partially reciprocal trade-offs in those components across frequency that coincide with Rabbitt's model results. However, we find that a simple kinetic model (meno presto, (Santos-Sacchi and Song, 2014b)) lacking piezoelectric coupling does the same. We also observed significant effects of initial state, characterized by initial $V_h$, on the frequency response of the reciprocal trade-offs. Furthermore, membrane tension differentially alters that frequency response depending on initial state. Whether such observations are in line with PZE theory remain to be determined.

**Methods**

All experimental protocols were approved by the Yale Animal Care and Use Committee and were in accordance with relevant guidelines and regulations. Detailed methods, including specifics of our voltage chirp stimulus protocol and extraction of complex NLC from patch admittance, are available in (Santos-Sacchi and Tan, 2019, 2020). Briefly, extracellular solution was (in mM): NaCl 100, TEA-Cl 20, CsCl 20, CoCl$_2$ 2, MgCl$_2$ 1, CaCl$_2$ 1, Hepes 10, pH 7.2. Experiments were performed at room temperature. Extracellular solution was in the patch pipette. On-cell macro-patches on the guinea pig OHC lateral membrane were made near the middle of the cylindrical cell, where prestin resides at high density (Dallos et al., 1991; Huang and Santos-Sacchi, 1993). Pipette inner tip size was about 3.5 μm, series resistance (R$_s$) estimated to be below 1 MΩ, and seals at 0 mV near 5 GΩ (see (Santos-Sacchi and Tan, 2020)). Under these conditions, R$_s$ effects are minimal (Santos-Sacchi and Tan, 2020) and were not corrected for in this report. Furthermore, in this report we make relative assessments of NLC among groups of macro-patches. Chirp voltage stimuli (10 mV pk) were superimposed on step holding potentials from -160 mV to +160 mV, in 40 mV increments. As detailed previously (Santos-Sacchi and Tan, 2020), the FFT derived admittance at +160 mV, where NLC is absent, is subtracted from admittance at all other step potentials, thereby removing stray capacitance. Additionally, because we focus on effects of V$_h$ on complex NLC behavior, we confirmed that subtractions for stray capacitance using either +160 or -160 mV responses give the exact same Boltzmann fits, as expected since stray capacitance will not influence prestin charge movement/NLC. A residual DC nonlinear conductance was also removed. DC conductance was determined from the DC component of FFT current response at each stepped voltage. Conductance, $\Delta I(0)/\Delta V_{hold}$, was gauged by spline interpolating between each step voltage response, and differentiating digitally. Since, such an approach can have untoward effects at the end point voltage extremes, we fit the conductance linearly between 0 and 396 Hz (conductance at each frequency determined from the real components), and used the resultant fitted conductance at zero Hz for subtraction of the real components of admittance across frequencies. **Fig. 2** illustrates, in a model, the necessity of such nonlinear DC conductance subtractions to validly extract the imaginary component of NLC. Patch membrane tension was imposed by changing pipette pressure. All data collection and analyses were performed with the software programs jClamp (www.scisoftco.com) and Matlab (www.mathworks.com). All means and standard errors (SE) are from individually analyzed patch data. Plots were made in Matlab.

In order to extract Boltzmann parameters, capacitance-voltage data were fit to the first derivative of a two-state Boltzmann function. We refer to this as the "C$_{sa}$ fit" in text and figures.

$$C_m = NLC + C_{sa} + C_{lin} = Q_{max} \frac{ze}{k_B T} \frac{b}{(1+b)^2} + C_{sa} + C_{lin} \qquad (m1)$$

$$\text{where} \quad b = exp\left(-ze\frac{V_m - V_h}{k_B T}\right), \ C_{sa} = \frac{\Delta C_{sa}}{(1+b^{-1})}$$

$Q_{max}$ is the maximum nonlinear charge moved, $V_h$ is voltage at peak capacitance or equivalently, at half-maximum charge transfer, $V_m$ is membrane potential, $z$ is valence, $C_{lin}$ is linear membrane capacitance, e is electron charge, $k_B$ is Boltzmann's constant, and T is absolute temperature. $C_{sa}$ is a component of capacitance that characterizes sigmoidal changes in specific membrane capacitance, with $\Delta C_{sa}$ referring to the maximal change at very negative voltages

(Santos-Sacchi and Navarrete, 2002; Santos-Sacchi and Song, 2014a).

For some data, a power fit as a function of frequency (*f*) was performed (Bai et al., 2019; Santos-Sacchi and Tan, 2019).

$$C(f) = C_0 + a * f^b \qquad\qquad (m2)$$

where $C_0$ is the asymptotic component, and *a* and *b* control the frequency response.

**Results**

Rabbitt (Rabbitt, 2020) modeled prestin and its NLC as a piezoelectric process based on first principles. NLC can be deconstructed into real and imaginary components, since it is defined as the frequency-dependent membrane admittance ($Y^*_m(\omega)$) divided by $i\omega$, where $\omega = 2\pi f$ and $i = \sqrt{-1}$ ,

$$C^*_m(\omega) = \frac{Y^*_m(\omega)}{i\omega} = \frac{B^*_m(\omega)}{\omega} - i\, \frac{G^*_m(\omega)}{\omega}$$

, providing a real capacitive component and an imaginary conductive component. On removing stray capacitance (by subtraction, see Methods) and linear capacitance (by Boltzmann fitting, see Methods) from $C^*_m(\omega)$, prestin-associated complex NLC($\omega$) is obtained.

With his PZE model, Rabbitt predicted that the imaginary component of complex NLC should increase over frequency, as the real component decreases. We previously reported on the real and imaginary components of complex NLC (Santos-Sacchi and Tan, 2020) and concluded that the imaginary component was small and essentially frequency independent. However, we did not fit both complex components to Boltzmann functions to unambiguously extract details of those components. In that paper, we focused our attention on the absolute magnitude of complex NLC, i.e., $\sqrt{Re(NLC)^2 + Im(NLC)^2}$, which is comparable to all previous measures of OHC/prestin NLC.

Here we explore in detail both components in 25 macro-patches of the OHC lateral membrane, first under zero pipette pressure, namely in the absence of externally applied membrane tension. For the 3D plots, black dots depict average data, and magenta dots depict standard error (+SE). The multicolored shading is provided by the *interp* surface plot function in Matlab, and the red and blue solid lines depict Boltzmann fits (see Methods) at selected frequencies.

**Figure 1** shows that indeed, Rabbitt's prediction concerning our data that he references is borne out. There appears to be a trade-off between real (**Fig.1 A**) and imaginary (**Fig.1 B**) component peak magnitude across frequency, especially revealed in the extracted 2-state fits of NLC (**Fig.1 C**). Superimposed lines in green and magenta are the PZE model responses at $V_h$ from Rabbitt (Rabbitt, 2020), which shows general agreement within this bandwidth. **Figure 1D** shows the absolute magnitude of NLC that continuously decreases across frequency. Thus, while there are reciprocal trade-offs between the real and imaginary NLC components, the trade-off is not fully reciprocal; that is, the overall kinetics of prestin (governed by a host of molecular loads) slows.

**Figure 2** (*top panel*) highlights the relationship between real and imaginary components of NLC. Here we plot extracted $Q_{max}$ from fitted NLC components across frequency. The trade-off in component magnitude shows that at a particular frequency ($F_{is}$), the two intersect, and above that frequency the imaginary component will dominate. Rabbit (Rabbitt, 2020) modeled that the imaginary component will eventually reduce to zero after peaking, but here (because of our limited frequency interrogation range) we use $F_{is}$ to explore influences on NLC frequency response (see below). **Figure 2** (*middle panel*) also shows that the voltage sensitivity ($z$) of the two components decreases in parallel across frequency, possibly indicating an increasingly impeded sensor charge movement within the membrane plane. In other words, voltage-induced conformational switching of prestin is depressed as frequency increases.

Another interesting observation that this fitting exercise exposes relates to the offset ($\Delta C_{sa}$) in capacitance at negative voltages, which is uniformly observed in all previous studies reporting on OHC

NLC. Indeed, when we first observed this offset and developed the method used to fit the response (Santos-Sacchi and Navarrete, 2002), we thought it was solely due to surface area alterations as prestin changed states from expanded to contracted. Surface area change in prestin embedded membrane was first reported by Kalinec et al. (Kalinec et al., 1992). However, the magnitude of $\Delta C_{sa}$ is greater than that expected for surface area changes alone in prestin. Thus, it likely includes changes to overall dielectric properties of the membrane, as well. **Figure 2** (*bottom panel*) compares the offsets in real and imaginary component fits by plotting the difference between the two. The magnitude of $\Delta C_{sa}$ is little affected across frequency, and, in fact, it can be seen in **Figure 1 A** and **B** that the offset is virtually absent in the imaginary component of NLC. Thus, the charge movement associated with the offset is solely capacitive in nature.

Do these trade-off observations necessarily indicate that prestin is piezoelectric? **Figure 3A** shows that the same basic behavior observed in our data is recapitulated by the *meno presto* kinetic model that we previously described (Santos-Sacchi and Song, 2014b). Both real and imaginary NLC components are observed, with trade-offs across frequency. Additionally, in this figure we illustrate the validity of our method of subtraction of a nonlinear DC conductance from our patch admittance data prior to subsequent analysis. With a nonlinear DC conductance included in the model (**Fig. 3B**), like that we find in membrane patches, the imaginary component is distorted. Following removal (**Fig. 3C**), complex NLC is equivalent to the model without such conductance (**Fig. 3A**). In our model, overall NLC behavior arises due to the stretched exponential transition rates between prestin's expanded and contracted states. There is no intrinsic piezoelectric coupling between voltage and tension embodied in this model. But, without the delays introduced in those conformational transitions (i.e., in a simple ultra-fast 2-state model), no imaginary component exists within our interrogation bandwidth. Thus, the requirement of piezoelectricity in prestin is not firmly established by the trade-off in real and imaginary components of NLC that we find across stimulus frequency.

The data we have presented thus far are from averaged NLC, but we know that NLC $V_h$ can vary among cells (and patches) due to a variety of reasons, e.g., membrane tension (Iwasa, 1993; Gale and Ashmore, 1994; Kakehata and Santos-Sacchi, 1995), anions (Oliver et al., 2001; Santos-Sacchi et al., 2006; Rybalchenko and Santos-Sacchi, 2008), temperature (Santos-Sacchi and Huang, 1998; Meltzer and Santos-Sacchi, 2001; Okunade and Santos-Sacchi, 2013) and initial voltage conditions (Santos-Sacchi et al., 1998; Santos-Sacchi et al., 2009). Consequently, we sought to characterize the dependence of NLC real and imaginary component behavior on prestin's initial state; namely, as typified by initial $V_h$, which reports on the distribution of proteins in either the expanded or contracted states. First, we categorized NLC data into groups possessing $V_h$ values above and below -30 mV.

**Figures 4** and **5** show these categorized averaged responses. Means of $V_h$ for the two groups are -48.8 +/- 3.4 mV (n=11) for the negative range group, and -17.5 +/- 2 mV (n=14) for the positive range group. Differences between the frequency response of the two groups are apparent, especially in imaginary components (**Fig. 5 A, B, C** and **Fig. 6 A, B, C**). These differences are highlighted by plotting the intersection of the fitted $Q_{max}$ magnitudes of each NLC component (**Fig. 5C and Fig. 6C**), where the intersection frequency ($F_{is}$) is at a lower frequency for the negative range group (16 kHz) than the positive range group (19.4 kHz). **Figure 6** recategorizes $V_h$ into 3 groups, and clearly shows a nonlinear increase of the intersection frequency with more depolarized initial $V_h$ conditions.

Membrane tension is well known to shift $V_h$ of OHC NLC. To investigate the influence of tension-induced $V_h$ shift on real and imaginary components of NLC, we altered the tension delivered to the macro-patch membrane (**Figure 7**). In 7 patches we were able to incrementally alter pipette pressure (thus, membrane tension) from 0 to - 2, -4, -6, -8, and -10 mm Hg (0 to – 1.22 kPa). Though substantial shifts in average $V_h$ occur (from -42.8 +/- 8 to -9.1 +/- 11.8 mV), only slight changes in the real or imaginary components of complex NLC are readily visible. However, upon plotting intersecting frequencies (*lower panels*), clear changes in $F_{is}$ are obvious. Increasing tension causes a shift in $F_{is}$ to higher frequencies. For 0 mmHg, $F_{is}$ is 12.9 kHz, whereas -10 mmHg pressure shifts $F_{is}$ to 28.3 kHz. Does initial $V_h$ factor into these responses?

As with our $V_h$ categorization observations made above, we sought to determine the influence of initial $V_h$ on tension effects. To this end, we categorized the patches into groups with initial $V_h$ above and below -30 mV. First, **Figure 8** illustrates that tension susceptibility depends on initial $V_h$, with the positive group (red symbols; 4.28 mV/mmHg) showing $V_h$ shift sensitivity nearly twice that of the negative group (blue symbols; 2.54 mV/mmHg). For all averaged patch responses (magenta symbols), the response was intermediate with standard errors the largest, as predicted from the categorization results.

In **Figure 9**, we plot average NLC for the negative initial $V_h$ group. **Figure 9A** shows results at 0 mmHg and **Figure 9B** shows results at -10 mmHg. Differences are apparent in the frequency responses, with **Figure 9C** highlighting the influence of increasing membrane tension on the intersection frequency, $F_{is}$. Increasing tension induces a shift in $F_{is}$ to higher frequencies. For 0 mmHg, $F_{is}$ is 11 kHz, whereas -10 mmHg pressure shifts $F_{is}$ to 16.2 kHz.

A similar analysis is shown in **Figure 10** for the positive initial $V_h$ group. Surprisingly, in this case, increasing tension induces a shift in $F_{is}$ to lower frequencies. The results for both groups are summarized in **Figure 11A**. This plot shows that the two initial $V_h$ group responses converge towards a common $F_{is}$ at -10 mmHg, namely about 16 kHz. It may not be a coincidence that this pressure value is close to the turgor pressure ($\sim$ 1 kPa) of the native OHC (Ratnanather et al., 1993). Finally, in **Figure 11B** we plot the absolute magnitude of NLC for the two initial $V_h$ groups. The negative group shows a very similar frequency response regardless of tension, and is in line with our previous observation on the immutable frequency response of the absolute magnitude of complex NLC during tension changes (Santos-Sacchi and Tan, 2020). However, the positive group shows altered frequency responses, with increases in tension slowing the frequency response. Our data thus indicate an interaction between initial prestin state and tension sensitivity.

**Discussion**

OHC NLC is frequency dependent and low pass. We first demonstrated this in guinea pig OHCs using AC voltage chirp stimuli to assess the real component of complex NLC nearly 30 years ago (Santos-Sacchi, 1991). Over the years this has been confirmed across species (Gale and Ashmore, 1997b; Albert et al., 2007; Homma et al., 2013; Santos-Sacchi and Tan, 2018). Indeed, combining measures across a number of studies, we arrived at a collective estimate of NLC (absolute magnitude) frequency response that led us to suggest that prestin activity could not drive eM to sufficiently influence cochlear mechanics at very high acoustic frequencies (>50 kHz), where cochlear amplification is expected to work best (Santos-Sacchi and Tan, 2020). Recently, Rabbitt (Rabbitt, 2020) has modelled OHC NLC as a piezoelectric process whose imaginary complex component takes on special significance, namely, signifying power output associated with prestin charge displacement. He modelled that as the real component of NLC decreases across frequency, the imaginary component increases. Thus, he suggested that dielectric loss in voltage-driven prestin charge movement is indicative of considerable influence at high frequencies, thereby overriding our suggestions that prestin is limited in its high frequency effectiveness (Santos-Sacchi and Tan, 2019, 2020).

Here we find that prestin's complex NLC displays a partially reciprocal trade-off between real and imaginary components across interrogating frequency that follows predictions based on the PZE model of prestin (Rabbitt, 2020). However, similar behavior is found with a simple kinetic model (*meno presto* model, (Song and Santos-Sacchi, 2013; Santos-Sacchi and Song, 2014b) that is not piezoelectric. Thus, this type of behavior is not necessarily due to piezoelectricity in prestin, and whether other predictions of the PZE model correspond to the observed prestin behavior that we find remains to be seen. For decades, we and others (Santos-Sacchi, 1990, 1991; Gale and Ashmore, 1997b; Iwasa, 1997; Santos-Sacchi and Tan, 2018) have suggested that NLC and eM behavior are governed by the conformational kinetics of OHC motor (prestin) transitions, in line with traditional biophysical concepts of voltage-dependent protein behavior. Indeed, when we initially realized that prestin kinetics were influenced by chloride ions (Song and Santos-Sacchi, 2013; Santos-Sacchi and Song, 2016), we suggested that delays introduced by stretched exponential transition rates in prestin would impact the phase of eM. Subsequently, we found that an eM phase lag (re voltage) develops across frequency and this could be influenced by chloride anions (Santos-Sacchi and Song, 2014c). It is likely that the frequency dependent imaginary component of NLC we observe here correlates with that out of phase mechanical response. Thus, we agree with Rabbitt (Rabbitt, 2020) that the imaginary component of NLC may have special significance in prestin function, regardless of whether prestin works as a piezoelectric device or not. Nevertheless, our data show that absolute magnitude of NLC (representing total charge, including in and out of phase components) decreases as a power function of frequency, being 40 dB down at 77 kHz (Santos-Sacchi and Tan, 2020). This is a minuscule fraction of voltage-driven prestin activity that exists at low frequencies. Based on PZE models (Iwasa, 2001; Dong et al., 2002; Rabbitt, 2020), mechanical loads can influence the frequency response of prestin. We agree that NLC and eM are susceptible to mechanical load, but the frequency response of whole-cell eM is slower than that of NLC, and likely corresponds to both cellular (external) and molecular (intrinsic) load components (Santos-Sacchi et al., 2019). Thus, with minimal influence of external loads, we view our patch data as providing the best estimate of prestin's intrinsic conformational switching limit (including all influential molecular impedances), as originally espoused by Gale and Ashmore (Gale and Ashmore, 1997b). Furthermore, by increasing pipette pressure, thus impeding global membrane patch movements, somewhat akin to imposing a high-impedance mechanical load to the OHC in microchamber experiments (Frank et al., 1999), we possibly reduce the influence of the pipette water column [see (Iwasa, 1997)] without major effects on absolute NLC frequency response.

Characteristics of OHC NLC have been known for some time to depend on initial conditions; for example, $V_h$ depends on initial holding potential and anion binding is state-dependent (Santos-Sacchi et al., 1998; Santos-Sacchi et al., 2001; Song and Santos-Sacchi, 2010). In an effort to determine whether complex NLC shows sensitivity to initial conditions, we categorized our data into classes that had initial $V_h$ values above and below -30 mV. We found effects of initial conditions on the frequency response of real and imaginary components, where a metric of this relationship, $F_{is}$, the intersection frequency of real and imaginary magnitude components varies with initial $V_h$. As initial $V_h$ shifts positively, $F_{is}$ increases nonlinearly (see **Fig. 6**).

We also found initial $V_h$ effects on the influence of static membrane tension on $F_{is}$. Whereas the group with initial $V_h$ positive to -30 mV shows a decrease in $F_{is}$ with increasing membrane tension, the opposite is found in the group with initial $V_h$ negative to -30 mV. At -10 mmHg (1.22 kPa), each group's $F_{is}$ intersects near 16 kHz (see **Fig. 11A**). Interestingly, this pressure value is close to the turgor pressure (about 1 kPa) of the native cylindrical OHC (Ratnanather et al., 1993).

Lastly, we found that the group with initial $V_h$ negative to -30 mV shows little variation in the frequency response of the absolute magnitude of NLC with membrane tension, similar to what we observed previously (Santos-Sacchi and Tan, 2020). However, the frequency response of the group with initial $V_h$ positive to -30 mV shows a decrease in the frequency response as membrane tension increases. From all these observations, it appears that the frequency response/magnitude of the imaginary component of NLC is mainly sensitive to initial $V_h$. Though adherence to a minimum phase system response might predict a complimentary relationship between real and imaginary components, viscoelastic non-minimum phase systems are known in biology (Recio-Spinoso et al., 2011; Kobayashi et al., 2020). Consequently, our data indicate that the state of prestin, characterized by initial $V_h$, influences frequency relationships between components of complex NLC, and thus, according to Rabbitt (Rabbitt, 2020), the power output of prestin activity. Nevertheless, changes in the frequency response are relatively small with our pertubations.

As we noted above, there are several factors that influence $V_h$. In the case of membrane tension, what could govern $V_h$? There are numerous descriptions of membrane proteins/channels that are influenced by membrane tension (Jin et al., 2020), with some possessing voltage-dependence; yet, interestingly, none have been deemed piezoelectric, per se. Rather, standard biophysical influences of load on a protein's conformational state are espoused. For example, some Kv channels are voltage and tension sensitive, and simple models where changes in the equilibrium constant for channel opening, or equivalently the ratio of forward to backward transition rates, can account for tension effects on channel activity (Schmidt et al., 2012). What other influences could there be in the absence of direct PZE effects or direct biophysical mechanical influences on prestin?

Several ideas come to mind. Perhaps the degree of cooperativity among prestin units alters with membrane tension. We recently provided evidence for negative cooperativity in prestin that was related to the density of prestin within the membrane (Zhai et al., 2020). Additionally, we previously modelled initial voltage influences on NLC as resulting from OHC molecular motor-motor (prestin) interactions (Santos-Sacchi et al., 1998). For Kv 7.4 channels, cooperativity has been found to impart mechanical sensitivity to the channel in OHCs (Perez-Flores et al., 2020). Another possibility is the presence of a conductive element influencing the interaction between voltage sensor charge and the membrane field, where that element alters with tension. Under this circumstance, the voltage sensed by the prestin charge may change during alterations in tension, leading to apparent $V_h$ shifts relative to voltage clamp commands. Could this be a viscoelastic coupled conductive element, thus providing time dependent changes in voltage sensed? We have observed multi-exponential time-dependent behavior in NLC at fixed voltages (Santos-Sacchi et al., 1998; Santos-Sacchi et al., 2009). Is $G_{metL}$, the tension-dependent prestin leakage conductance we have identified (Bai et al., 2017), the conductive component? We are currently testing complex NLC in prestin mutants that have reduced $G_{metL}$ conductance.

Finally, the shift in $V_h$ could additionally result from changes in the membrane surface charge or dipole potential that may accompany changes in membrane tension. Warshaviak et al. (Warshaviak et al., 2011) have found that physiologically relevant changes in membrane tension could shift that potential by tens of millivolts. Such effects could produce changes in the distribution of voltage-dependent protein conformations that would be evident as altered $V_h$, despite effective voltage clamp. Interestingly, changes in temperature could conceivably alter the membrane dipole (in addition to many other properties of the membrane), since electrical breakdown in membrane is temperature dependent (Gneno et al., 1987). Thus, our observation of shifts in $V_h$ due to temperature (Santos-Sacchi and Huang, 1998; Meltzer and Santos-Sacchi, 2001; Okunade and Santos-Sacchi, 2013) could also be independent of direct action on prestin. Of course, in this case of these "charge screening" effects, alterations in transition rates would not underlie $V_h$ shifts, so this is unlikely to fully account for prestin's tension responsiveness since we do find changes in the frequency response of real and imaginary components of complex NLC. Hence, transition rates are indeed altered.

In sum, we find that prestin's NLC displays partial reciprocal trade-offs in magnitude across frequency. The trade-off is prestin state-dependent in that the frequency response of the components alter with initial $V_h$ and membrane tension. However, these observations do not nullify our initial observations (Santos-Sacchi and Tan, 2019, 2020) that absolute complex NLC, signifying total charge moved, decreases precipitously as a power function of frequency; thus, prestin charge displacement that correlates with electromotility (Santos-Sacchi and Tan, 2018) is expected to have limited physiological influence at very high frequencies.

**Legends**

**Figure 1** Real and imaginary components of patch NLC. **A)** Real component of complex NLC across interrogating frequency. Note continuous decrease as frequency increases. Fits are made to eq.1 to extract the 2-state (blue line) response at selected frequencies. $V_h$ is average between 0.244 and 2.44 kHz. **B)** Imaginary component of complex NLC. In this case, magnitude increases with frequency, but there is not a full reciprocal trade-off with the real component. **C)** Plot of the 2-state fits across frequency. Superimposed is the PZE model response at $V_h$ of Rabbitt (2020), which shows general agreement within this bandwidth. **D)** Plot of the absolute magnitude of NLC, namely, $\sqrt{Re(NLC)^2 + Im(NLC)^2}$ , indicating a continuous reduction of prestin activity across frequency.

**Figure 2** Frequency dependence of complex NLC components. ***Top panel)*** Plot of fitted $Q_{max}$ of 2-state components of complex NLC. The real and imaginary component magnitudes intersect at a particular frequency, $F_{is}$. Here the imaginary component absolute magnitude is plotted. Fits (solid lines) are with eq. 2. Red is real component; blue is imaginary component. $F_{is}$ is 18.5 kHz. ***Middle panel)*** Plot of the Boltzmann parameter, *z*. Both decrease in parallel with frequency, with the imaginary component being smaller, indicating a shallower voltage sensitivity. ***Bottom panel)*** Plot of difference between real and imaginary $\Delta C_{sa}$, showing some low frequency dependence. **Fig. 1 A** and **B** indicate that $\Delta C_{sa}$ is largely absent in the imaginary component.

**Figure 3** Real and imaginary components of meno *presto model* NLC. **A)** Responses are similar to patch data in Fig. 1. **B)** Same model as in **A**, but with an additional nonlinear DC conductance (designed to be like that in our patch data) which distorts the low frequency response of the imaginary component. **C)** After removing the DC conductance with the approach detailed in the **Methods** section, the response is essentially the same as in **A**, where no DC conductance is included, confirming the validity of our approach to remove any residual DC conductance in our patch data.

**Figure 4** Complex NLC patch data for patches categorized with $V_h$ more negative than -30 mV. **A, B, C)** Legends to **Figs. 1 A, B, C** analogously apply here. **D)** Plot of fitted $Q_{max}$ of 2-state components of complex NLC. The real and imaginary component magnitudes intersect at a particular frequency, $F_{is}$. Here the imaginary component absolute magnitude is plotted. Fits (solid lines) are with eq. 2. Red is real component; blue is imaginary component. $F_{is}$ is 16 kHz.

**Figure 5** Complex NLC patch data for patches categorized with $V_h$ more positive than -30 mV. **A, B, C)** Again, legends to **Figs. 1 A, B, C** analogously apply here. **D)** Plot of fitted $Q_{max}$ of 2-state components of complex NLC. The real and imaginary component magnitudes intersect at a particular frequency, $F_{is}$. Here the imaginary component absolute magnitude is plotted. Fits (solid lines) are with eq. 2. Red is real component; blue is imaginary component. $F_{is}$ is 19.4 kHz.

**Figure 6** $F_{is}$ for complex NLC patch data for patches recategorized into 3 $V_h$ groups. Error bars are SE. Note that as $V_h$ is more negative, $F_{is}$ increases nonlinearly. This shows that $V_h$ influences the frequency responses of real and imaginary components of complex NLC.

**Figure 7** Effects of membrane tension on complex NLC. Pipette pressure was set to **A)** 0 mmHg, **B)** -4 mmHg, **C)** -8 mmHg, and **D)** -10 mmHg. Responses are characterized as in **Fig. 1** legend. The *bottom panels* show that $F_{is}$ increases as tension increases.

**Figure 8** Shift in $V_h$ due to tension versus initial $V_h$. This plot shows that sensitivity to tension differs depending on initial $V_h$. Upon categorization of initial $V_h$ with values greater than and less than -30 mV, differences in sensitivity to tension are exposed. In those patches with initial $V_h$ greater than -30 mV (red symbols; -23.0 +/- 2.6 mV; 4.28 mV/mmHg; n=3), sensitivity is nearly double that of the group with initial $V_h$ less than -30 (blue symbols; -57.6 +/- 5.4 mV; 2.536 mV/mmHg; n=4). For comparison, the relationship for all average patches is shown (magenta symbols), where a sensitivity to tensions is 3.3 mV/mmHg. This latter relationship is derived from data in **Fig. 7**. Linear fits provide the slope sensitivity.

**Figure 9** Membrane tension effects on the frequency response of complex capacitance for negative initial $V_h$. **A)** 0 mmHg, **B)** -10 mmHg. Clear changes, especially in the imaginary component, are noted between the two tensions. These changes are highlighted in **C)** where the intersection frequency, $F_{is}$, is found to *increase* with increases in tension. A shift from 11 to 16.2 kHz occurs between the two extremes.

**Figure 10** Membrane tension effects on the frequency response of complex capacitance for positive initial $V_h$. **A)** 0 mmHg, **B)** -10 mmHg. Clear changes, especially in the imaginary component, are noted between the two tensions. These changes are highlighted in **C)** where the intersection frequency, $F_{is}$, is found to *decrease* with increases in tension. A shift from 22.1 to 16.8 kHz occurs between the two extremes.

**Figure 11** Summary of initial $V_h$ influence on tension effects. **A)** The direction of change in $F_{is}$ is dependent on initial $V_h$. Here analyses were made with stray capacitance removal using either the +160 mV (circles) or -160 mV (asterisks) holding admittance. Results are identical, as expected. The two initial $V_h$ group responses converge until a common $F_{is}$ vs. tension is reached at -10 mmHg, namely about 16 kHz. **B)** Plot of the absolute magnitude of NLC for the two initial $V_h$ groups. The negative group shows a very similar frequency response regardless of tension, but the positive group shows altered frequency responses, with increases in tension slowing the frequency response.

Fig. 1

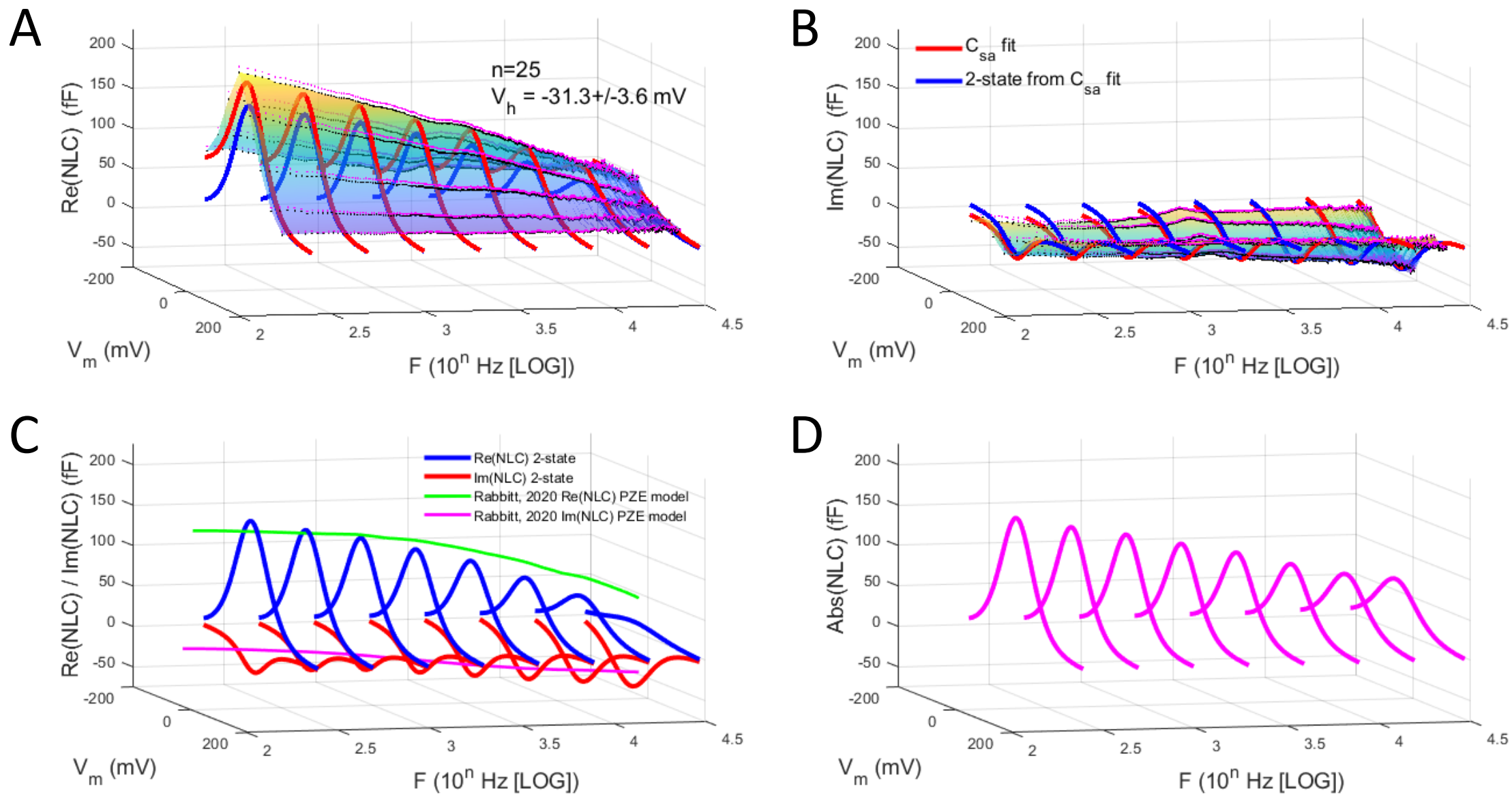

State dependence of complex NLC Santos-Sacchi et al.

Fig. 2

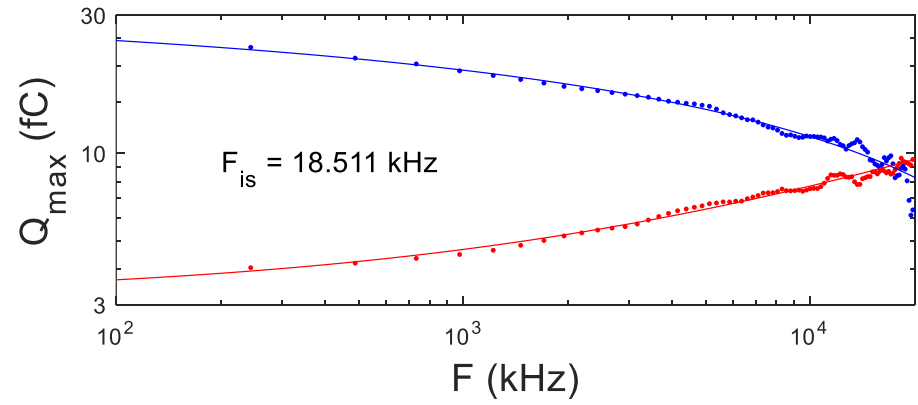

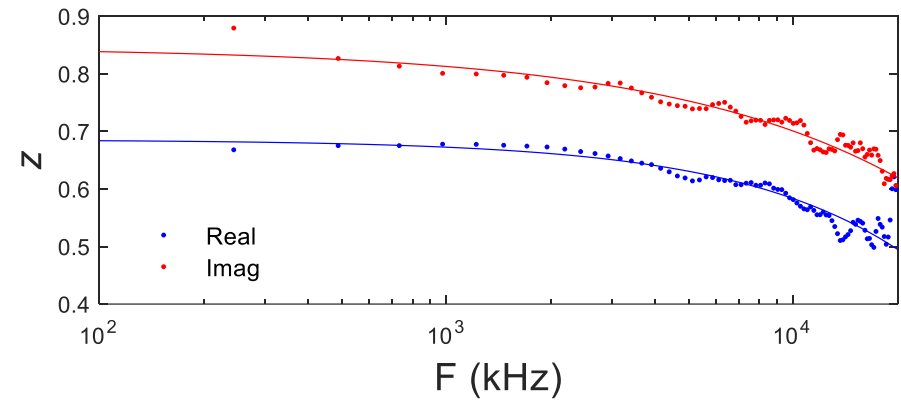

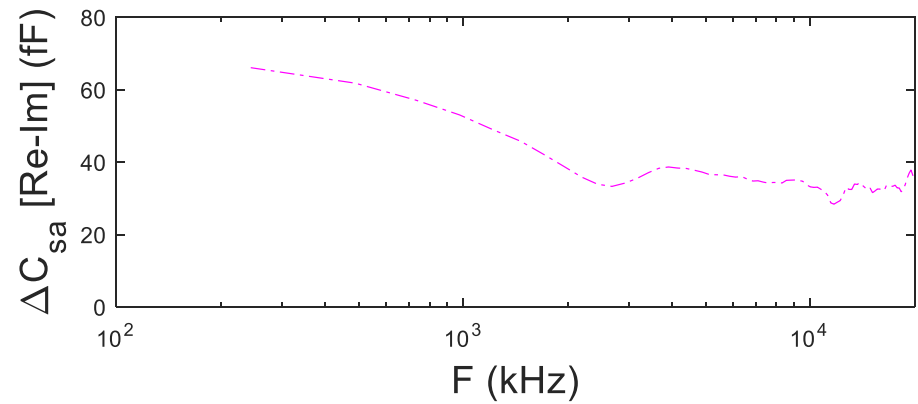

State dependence of complex NLC Santos-Sacchi et al.

Fig. 3

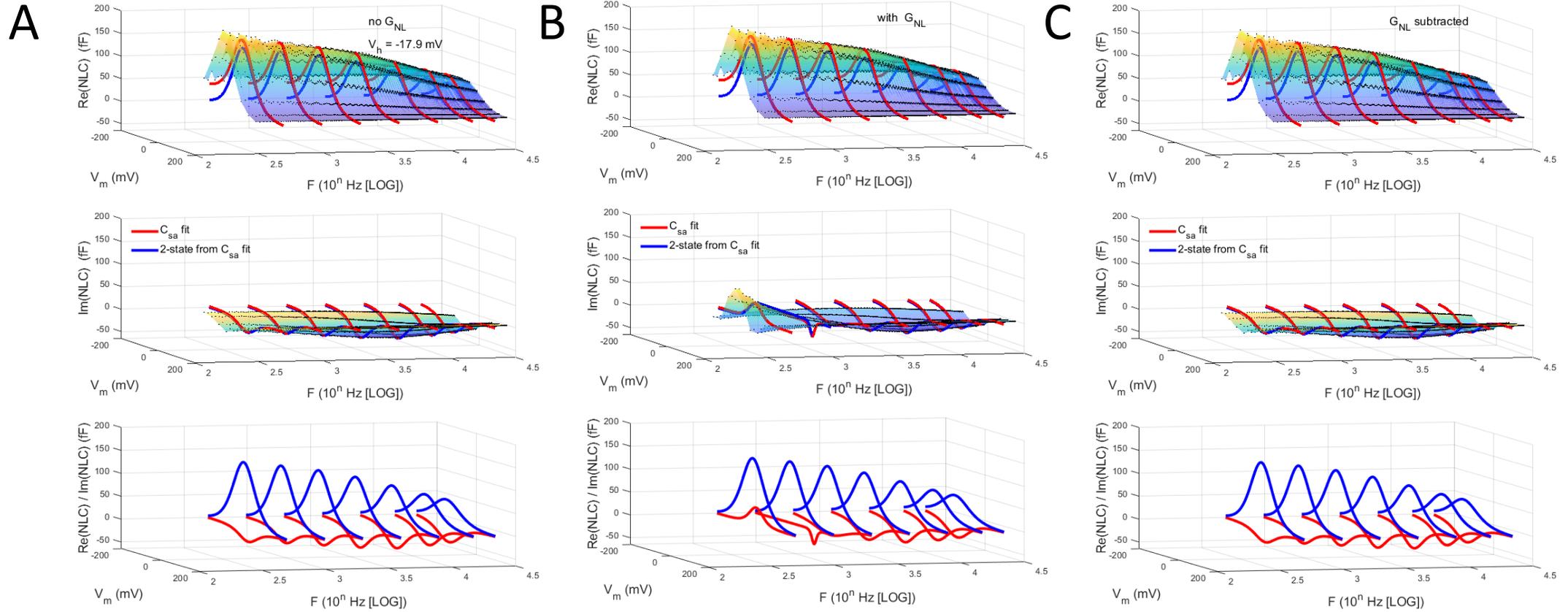

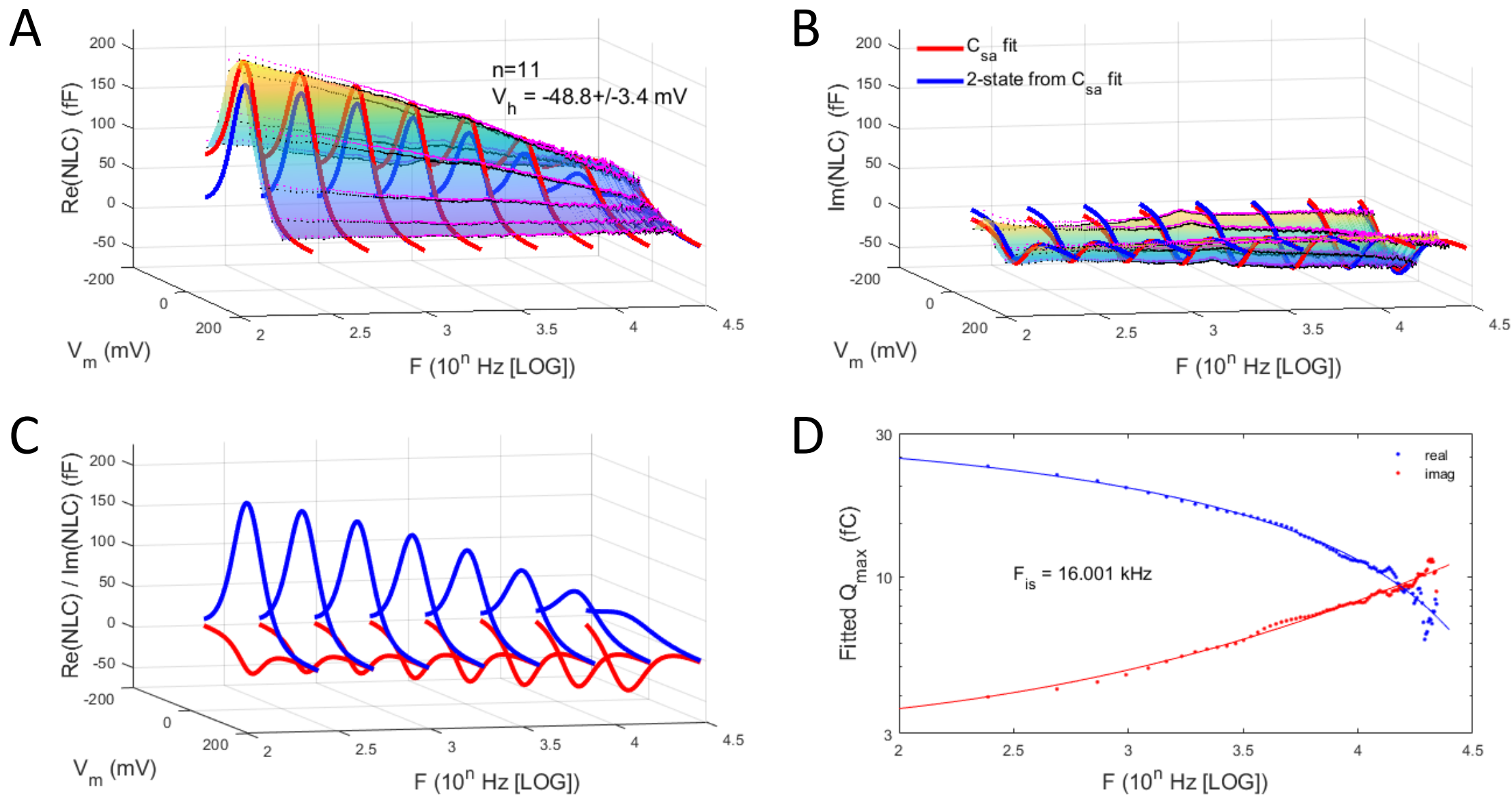

Fig. 5

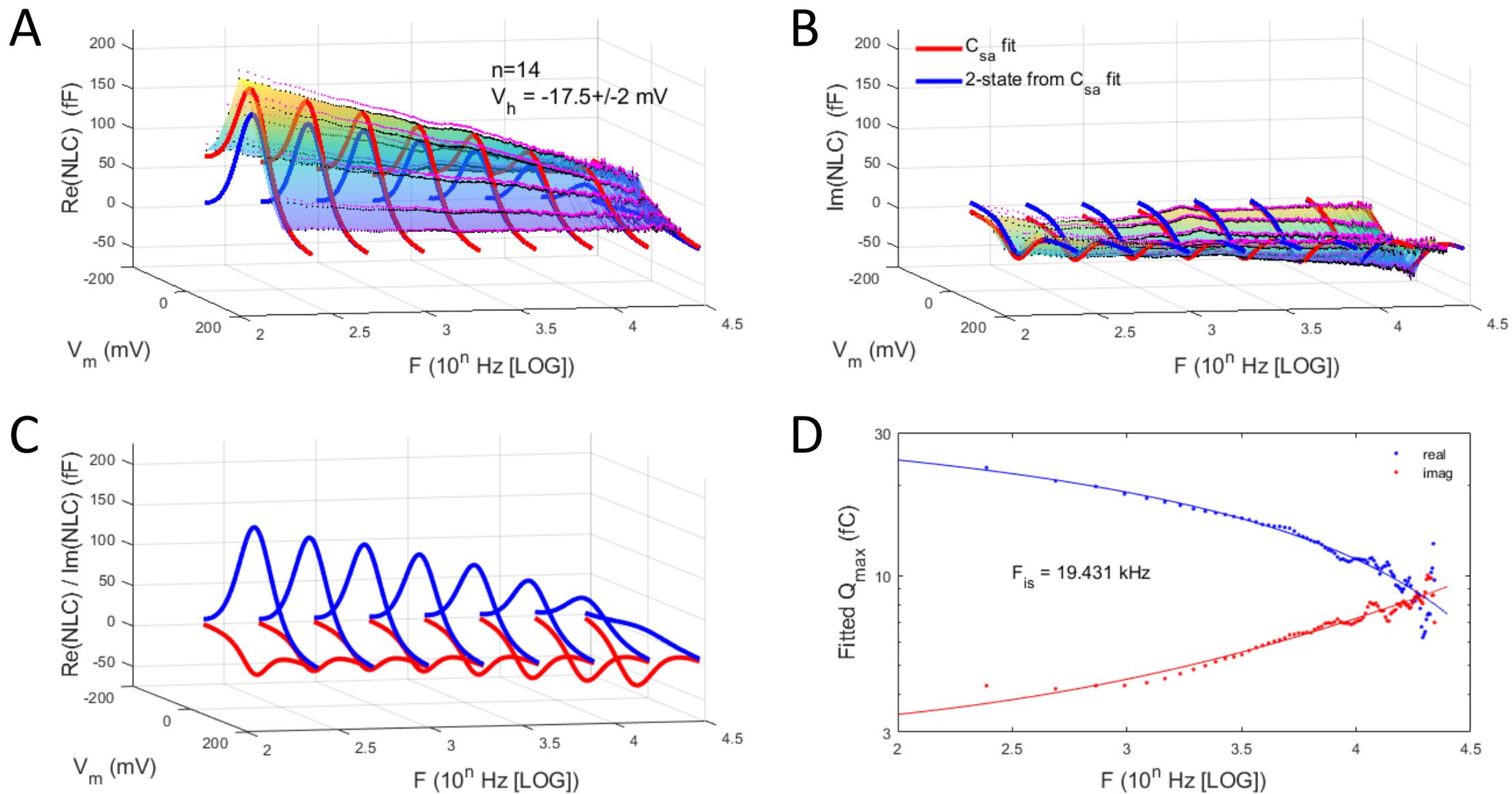

State dependence of complex NLC Santos-Sacchi et al.

Fig. 6

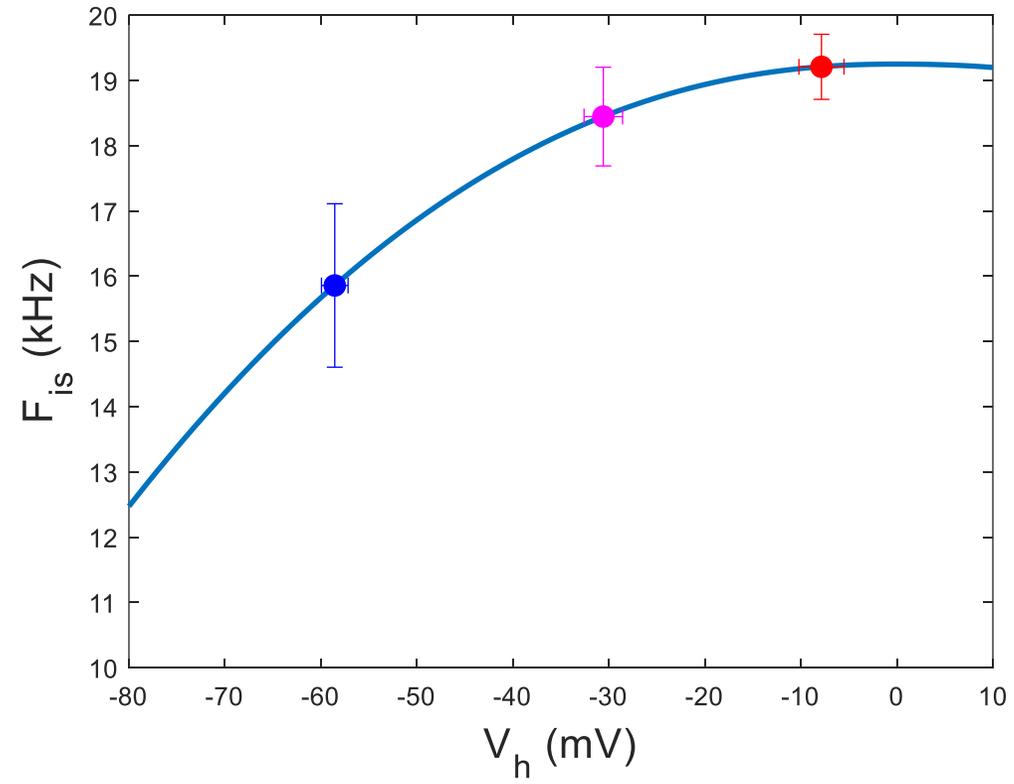

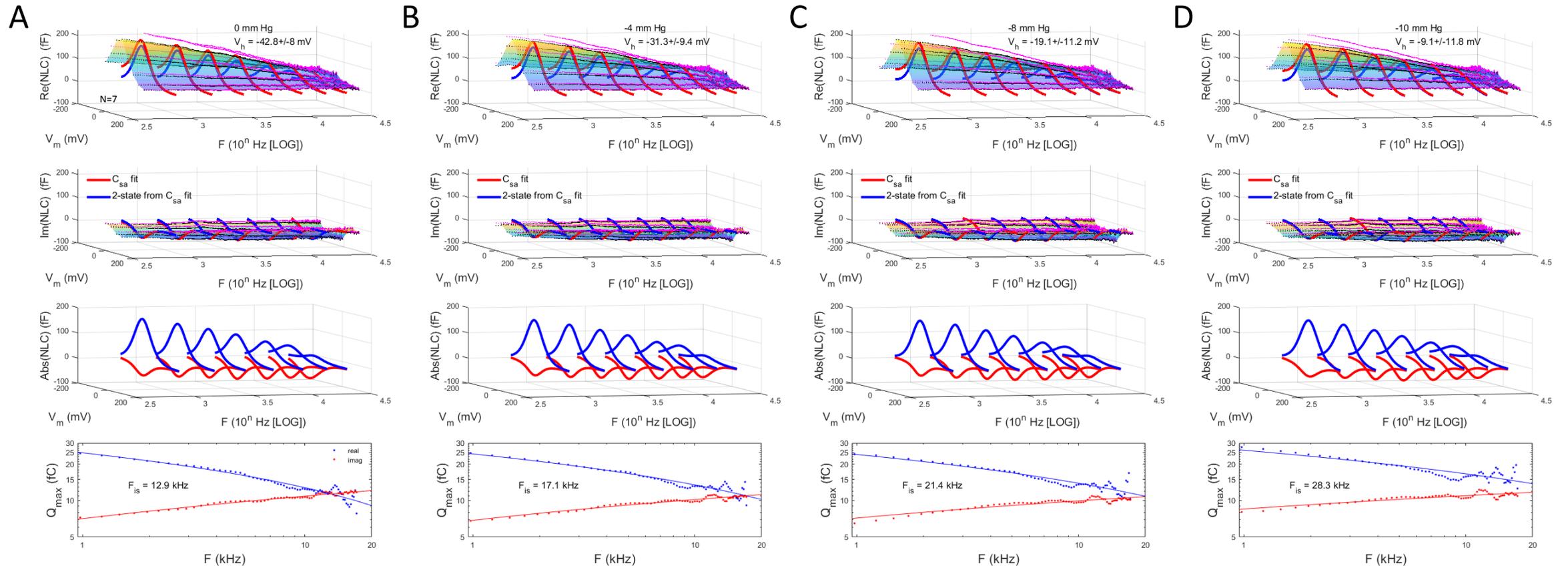

Fig. 8

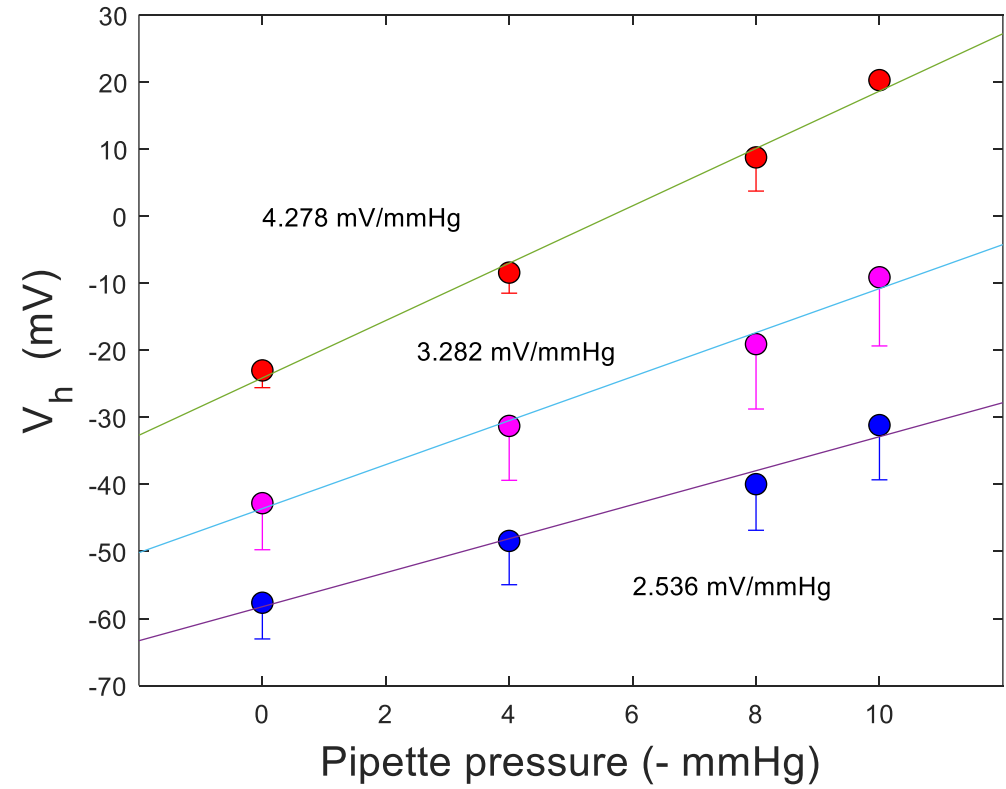

Fig. 9

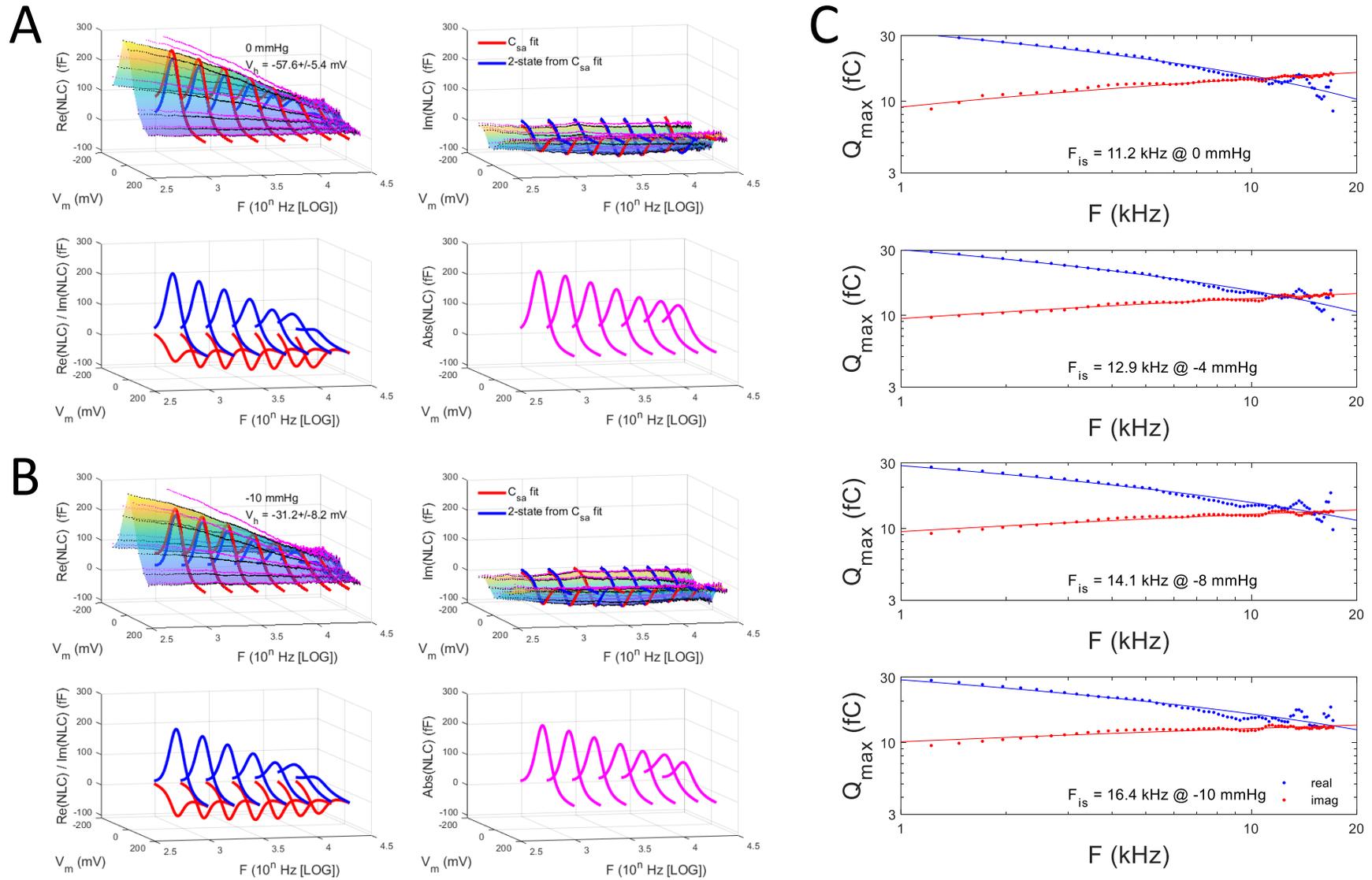

State dependence of complex NLC Santos-Sacchi et al.

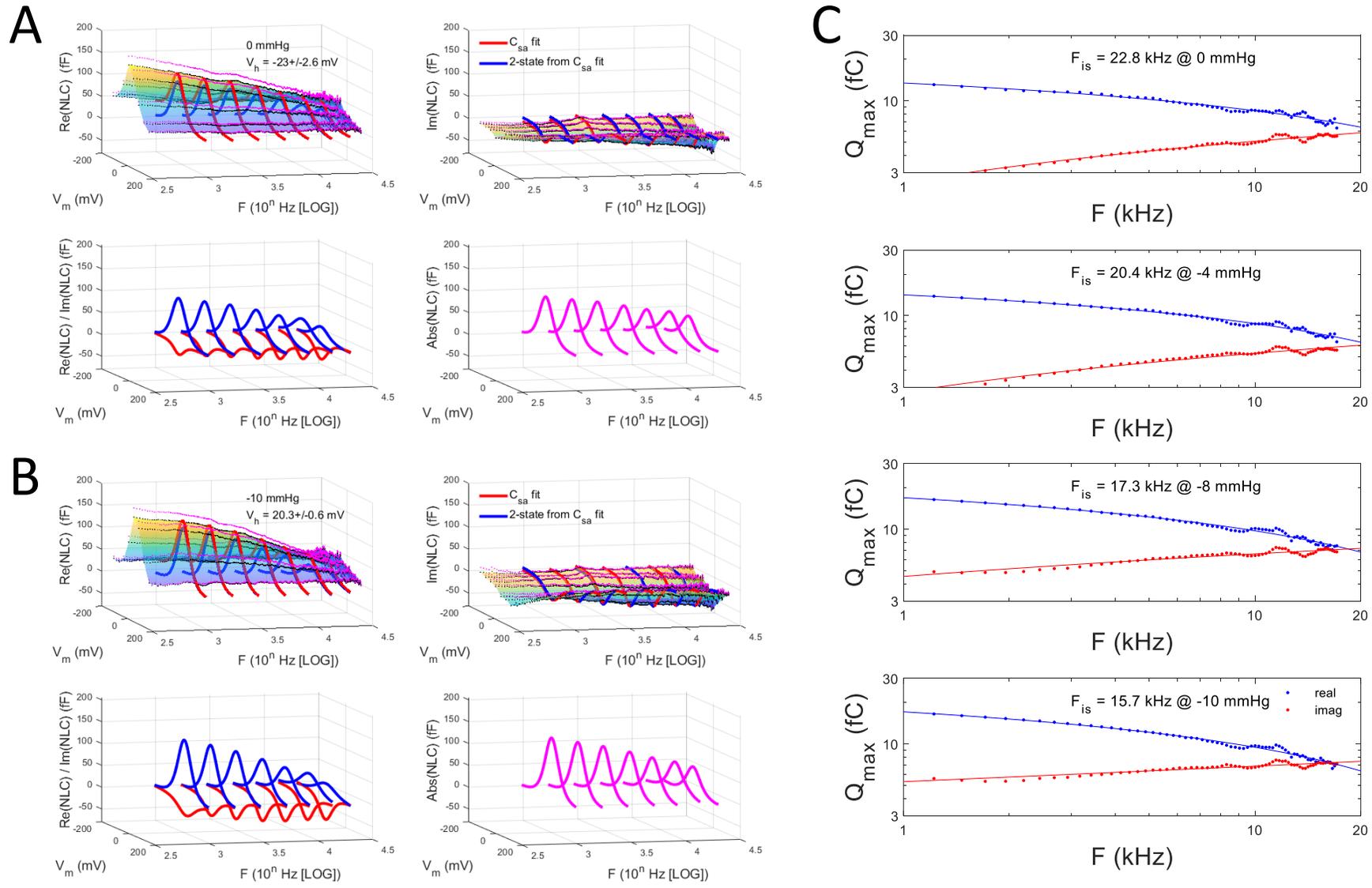

Fig. 11

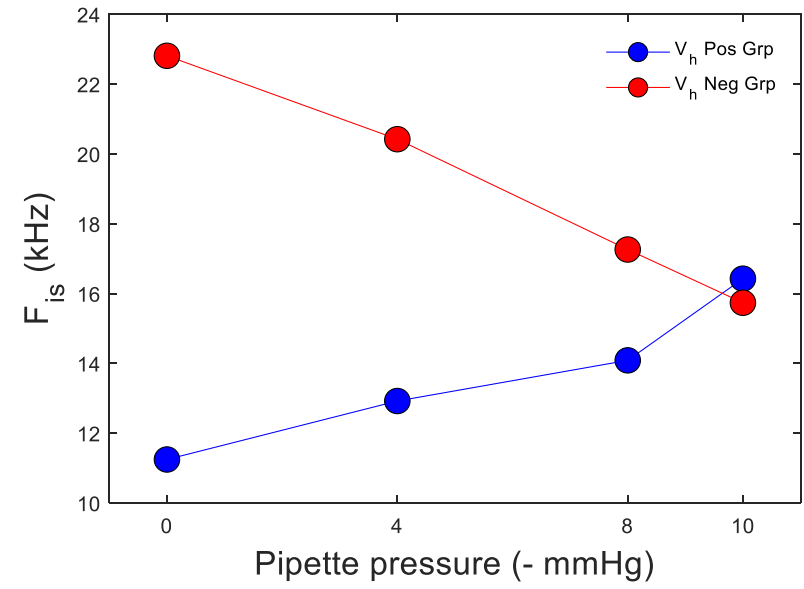

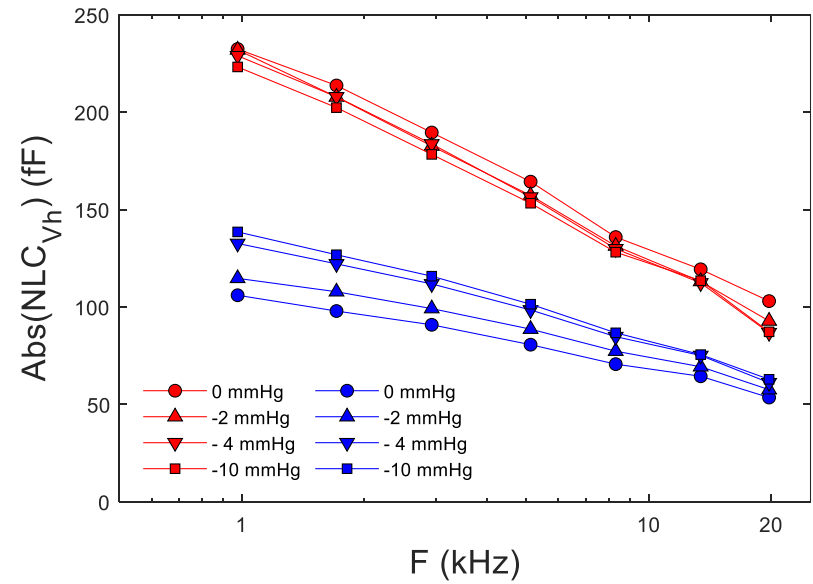

State dependence of complex NLC Santos-Sacchi et al.